# Tunneling spectroscopy of persistent currents in superconducting microrings

K. Yu. Arutyunov[a*], T. T. Hongisto[a], D. Y. Vodolazov[b]

[a]*University of Jyväskylä, Department of Physics, PB 35,*

*40014 Jyväskylä, Finland*

[b]*Institute for Physics of Microstructures, Russian Academy of*

*Sciences, 603950, Nizhny Novgorod, GSP-105, Russia*

**Abstract**

It is shown that in a structure consisting of a superconducting ring-shaped electrode overlapped by a normal metal contact through a thin oxide barrier, measurements of the tunnel current in magnetic field can probe persistent currents in the ring. The effect manifests itself as periodic oscillations of the tunnel current through the junction at a fixed bias voltage as function of perpendicular magnetic field. The magnitude of oscillations depends on bias point. It reaches maximum at energy $eV$ which is close to the superconducting gap and decreases with increase of temperature. The period of oscillations $\Delta\Phi$ in units of magnetic flux is equal neither to $h/e$ nor to $h/2e$, but significantly exceeds these values for larger loop circumferences. The phenomenon is explained by formation of metastable states with large vorticity. The pairing potential and the superconducting density of states are periodically modulated by the persistent currents at sub-critical values resulting in corresponding variations of the measured tunnel current.

## 1. Introduction

Diamagnetic response (Meissner effect) is a fundamental property of a superconductor. Periodic boundary conditions result in a very peculiar behavior in external magnetic field, e.g. flux quantization [1,2]. When a thin-wall superconducting ring is exposed to a perpendicular magnetic field persistent current circulate along the perimeter. The energy associated with these current states $E_L$ is given [3] by:

$$E_L \sim \frac{1}{S}\left(\frac{\Phi}{\phi_0} + L\right)^2, \quad (1)$$

where the winding number ('vorticity') $L = \frac{1}{2\pi}\oint \nabla\theta ds$ is an integer, $\theta$ is the coordinate-dependent phase of the superconducting order parameter $\Delta = |\Delta|e^{i\theta}$. Integration is made along the contour of the loop with circumference $S$. $\Phi$ is the magnetic flux through the area of the loop, and $\phi_0 = h/2e$ is the superconducting flux quantum. If the system can relax to its ground state, then magnetic field sweep causes periodic variation of kinetic properties with the period $\Delta\Phi = \phi_0$, corresponding to transitions $\Delta L = +/-1$ (Fig. 1 top panel, solid thick line). Persistent current in the loop is proportional to the derivative of the energy $I \sim dE/d\Phi$, and shows the characteristic saw-tooth behavior with the same period (Fig. 1 bottom panel, solid line). Switching from a state with vorticity $L$ to the nearest state $L+/-1$ happens at supercurrent density corresponding to accumulation of the phase $\Delta\theta$ equal to $2\pi$ along the perimeter of the loop: $j^0_{switch} \sim 2\pi / S$. The $\Delta\Phi = \phi_0$ periodicity is commonly observed at temperatures close to the critical one [4] when the superconducting energy gap $\Delta(T)$ is much smaller than the thermal energy $k_BT$.

---

[*] Corresponding author. Tel.:+358-14-260-2609; fax: +358-14-260-2351; e-mail:Konstantin.Arutyunov@phys.jyu.fi.

However, at low temperatures $k_BT<<\Delta(T)$ the system may not relax to the ground state being 'frozen' at a metastable state (Fig. 1, dashed lines). The ultimate condition of changing the winding number $L$ is the equivalence of the magnitude of the persistent current to the critical value. The critical current is reached when a phase difference $2\pi$ is accumulated on the superconducting coherence length: $j_c \sim 2\pi/\xi$. Thus, in loops with perimeter $S>>\xi$ one may observe vorticity changes $\Delta L \sim j_c / j^0_{switch} \sim S / \xi >> 1$. In recent experiments [5,6] at temperatures $T \sim T_C / 3$ measurements of magnetization of small superconducting rings did show transitions with changes of vorticity $\Delta L$ few times larger than unity.

## 2. The model

In the present work we analyze tunneling experiments probing persistent currents in superconducting loops at ultra-low temperatures $k_BT<<\Delta(T)$. The system under investigation is a normal metal – insulator - superconductor (NIS) structure, where the superconducting electrode has the shape of a loop (Fig. 2). It has been found [7] that the tunnel current at a fixed voltage bias periodically oscillates in perpendicular magnetic field (Fig. 3). The explanation has been provided [8] based on observation that the tunnel current of a NIS junction depends on the superconducting density of states, which is periodically modulated by the persistent current at sub-critical values [9,10]. The model [8] gives reasonable agreement with experiment for the period of tunnel current oscillations. However model [8] cannot explain the maximum in the dependence of the normalized magnitude of current oscillations $\Delta I/I_{max}$ on the bias voltage close to $eV/\Delta \sim 1$ (Ref. [7], Fig. 5). In this report we extend our previous model [8] including the contribution of sub-gap currents. To account for this process we use 'classical' BTK model [11] for pure superconductors and the earlier model [8]. This leads to the semi-quantitative result for the sub-gap current:

$$I_{sub} = \alpha |T|^2 \int_{-\infty}^{+\infty} A(\Delta,E)[f(E, T) - f(E + eV, T)]dE \quad (2)$$

where the probability of Andreev reflection $A(\Delta,E)$ is taken from Table II of Ref. [11]:

$$A(\Delta,E) = \frac{\Delta^2}{E^2 + (\Delta^2 - E^2)(1+2Z^2)^2}, \quad E < \Delta, (3a)$$

$$A(\Delta,E) = \frac{\Delta^2}{\left(E^2 - \Delta^2\right)\left(\sqrt{E^2/(\Delta^2 - E^2)} + 1 + 2Z^2\right)^2}, \quad E > \Delta, (3b)$$

where $Z$ is the dimensionless barrier strength, and $\Delta$ is found from the corresponding Usadel equations [8]. The tunnel current $I_{tun}$ and the sub-gap current $I_{sub}$ have different functional dependencies on $\Delta$, and hence their variation in magnetic field is also different. While sweeping the magnetic field the magnitude of the oscillations of the tunnel current is much higher than the magnitude of oscillations of the sub-gap current. Strictly speaking, this over-simplified introduction of a sub-gap current is not applicable for dirty limit electrodes studied in Ref. [7]. However, it demonstrates that at energies $\Delta << eV$ an account for the sub-gap current contribution gives qualitatively better agreement with experiment: Fig.4. The described approach is not able to give quantitative exact values for the sub-gap current as other mechanisms (e.g. leakage current) might contribute to the total current measured in experiment. It is also known that the magnitude of a sub-gap current strongly depends on interference effects within the locus of the NIS junction, and hence is geometry-dependent [12].

In spite of obvious simplifications describing the sub-gap current, the proposed approach is able to give good qualitative and reasonable quantitative agreement with experiment [7]. In a full agreement with these experimental findings, the calculated value of the magnitude of current oscillations (Fig. 3) and the normalized magnitude $\Delta I/I_{max}(B, V = const)$ (Fig. 4) depend on the bias voltage $V$. The only fitting parameter is the barrier strength $Z$. At high temperatures the current across a tunnel junction at $eV << \Delta$ is mainly determined by the tunnel component due to the temperature smearing of the Fermi distribution function. While at very low temperatures in the same limit the contribution of tunnel current is negligible, and the finite measured current is practically equal to the sub-gap term $I_{sub}$. For high biases $eV > \Delta$ at all temperatures the total current is determined by the tunnel component $I_{tun}$. Due to relatively weak dependence of the sub-gap current on magnetic field, dependence of $\Delta I/I_{max}$ versus voltage shows pronounced maximum at low temperatures, and monotonous behavior at higher temperatures (Fig. 4).

### 3. Conclusions

Using microscopic approach we have analyzed ultra-low temperature behavior of a mesoscopic-size NIS junction with a loop-shaped superconducting electrode [7]. It has been already shown [8] that in such systems the tunnel current oscillates in magnetic field with a period, which scales with the loop circumference. For large loops flux changes are much larger than the flux quantum. In present work we extend the approach [8] considering contribution of sub-gap currents. Using simple BTK model, one is able to obtain a reasonable quantitative agreement with the experimental results of Ref. [7].

### Acknowledgements

The work was supported by the EU Commission FP6 NMP-3 project 505587-1 SFINX.

Fig. 1. Dependence of energy (top) and persistent current (bottom) on the magnetic flux for a superconducting loop. Solid lines stand for 'conventional' case with $h/2e$ periodicity. Dashed lines correspond to formation of metastable energy states resulting in higher than $h/2e$ periodicity.

Fig. 2. Schematic of the system under consideration.

Fig. 3. Magnetic field dependence of the total current through a NIS junction with loop-shaped superconducting electrode at different values of the bias voltage. Parameters are: $T_c = 1.2$ K, $\xi(0) = 150$ nm, side of the square loop $w = 5$ μm, line width $d = 120$ nm, $T = 106$ mK.

Fig. 4. Magnetic field dependence of the normalized magnitudes of the low field (B < 5 mT) tunnel current oscillations $\Delta I/I_{max}$ as the function of the normalized bias $eV/\Delta(T)$ for the same structure as in Fig.2. $I_{max}$ is defined as the maximum tunnel current before switching to a new quantum state.

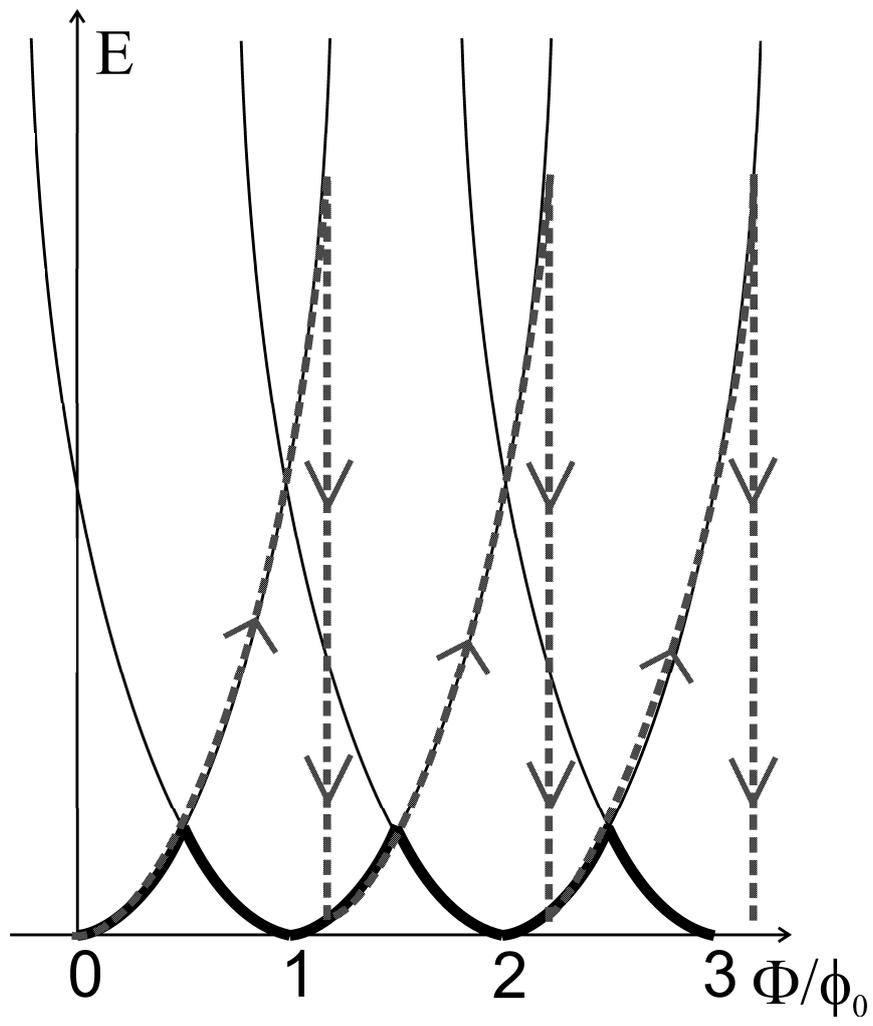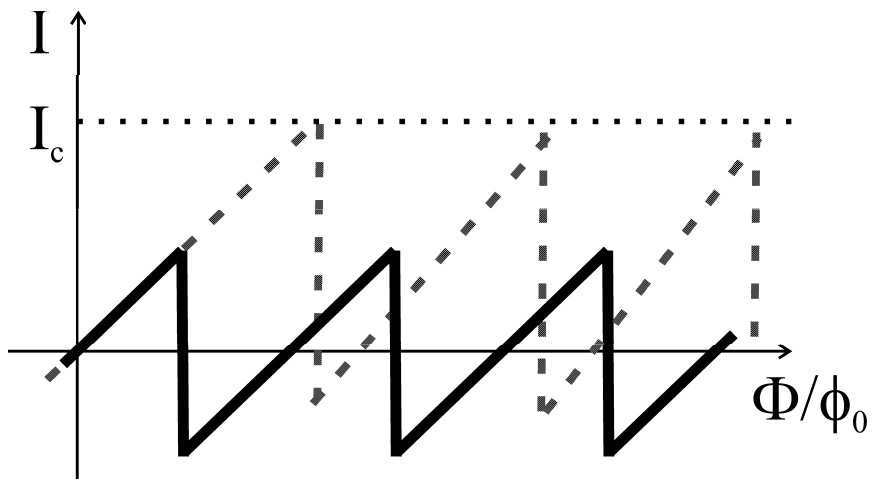

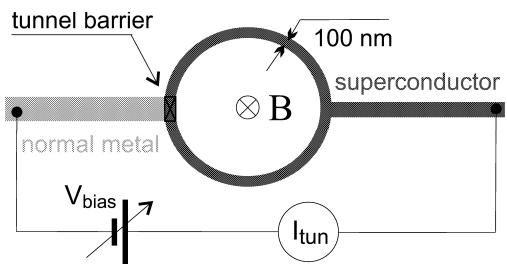

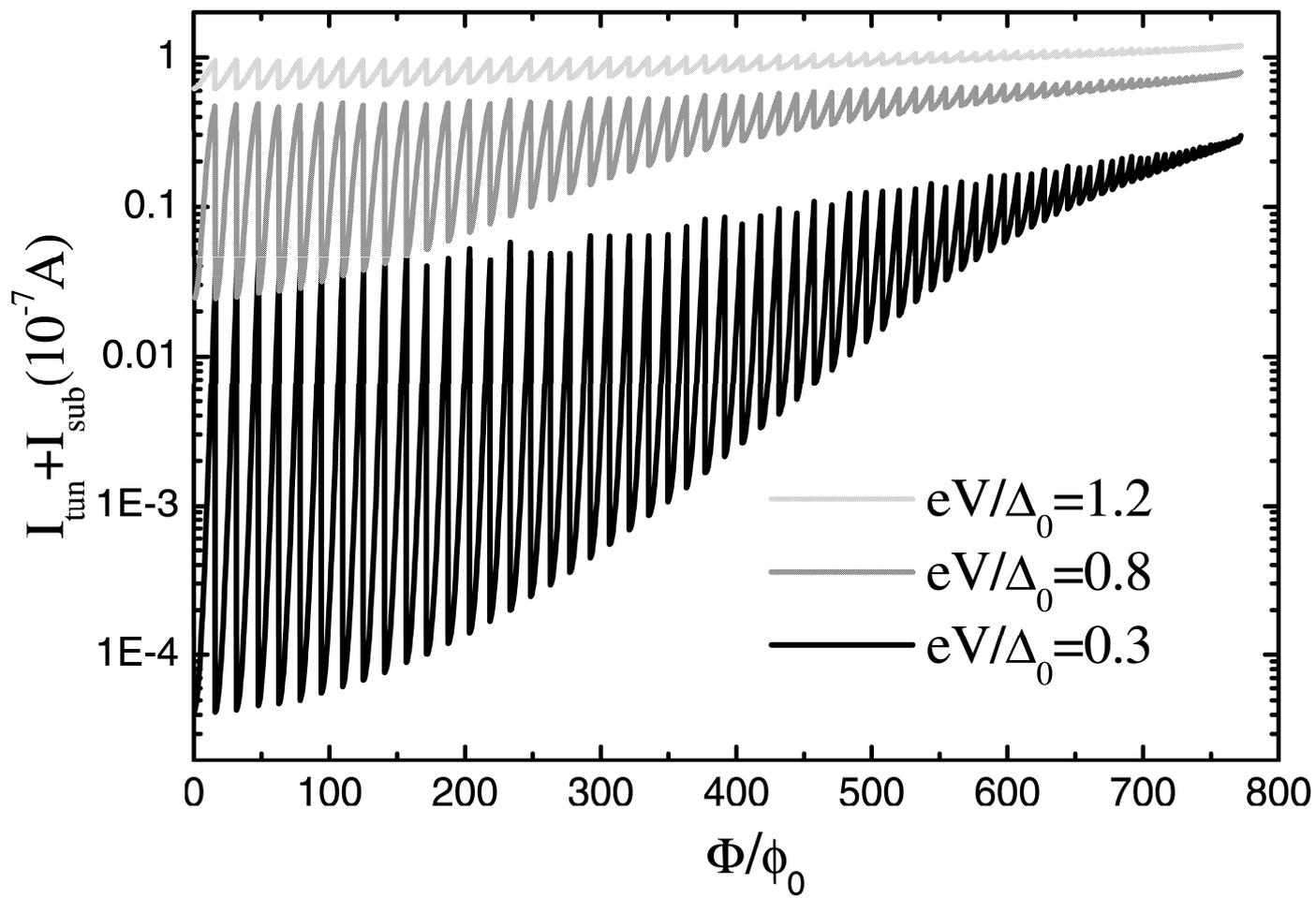

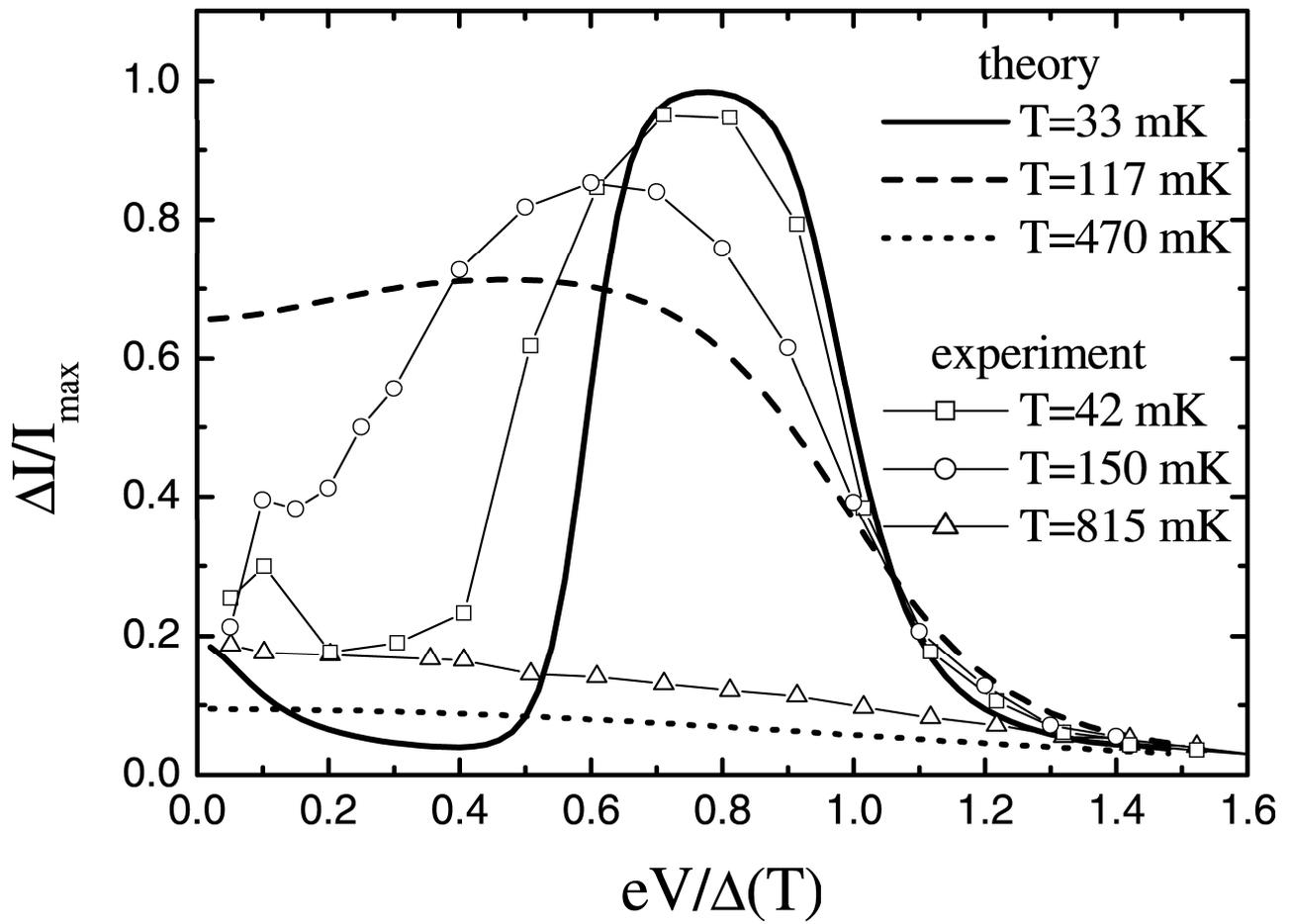